# Fast Adaptive Beamforming based on kernel method under Small Sample Support

Hu Xie[*], Da-Zheng Feng and Ming-Dong Yuan

National Lab. of Radar Signal Processing, Xidian University, Xi'an, 710071, China

**Abstract:**

It is well-known that the high computational complexity and the insufficient samples in large-scale array signal processing restrict the real-world applications of the conventional full-dimensional adaptive beamforming (sample matrix inversion) algorithms. In this paper, we propose a computationally efficient and fast adaptive beamforming algorithm under small sample support. The proposed method is implemented by formulating the adaptive weight vector as a linear combination of training samples plus a signal steering vector, on the basis of the fact that the adaptive weight vector lies in the signal-plus-interference subspace. Consequently, by using the well-known linear kernel methods with very good small-sample performance, only a low-dimension combination vector needs to be computed instead of the high-dimension adaptive weight vector itself, which remarkably reduces the degree of freedom and the computational complexity. Experimental results validate the good performance and the computational effectiveness of the proposed methods for small samples.

## I.  Introduction:

Wide applications of the large-scale adaptive beamforming (BF) have been found in radar [1], sonar [2], seismology [3], radio astronomy [4], microphone array speech processing [5], and wireless communications [6] and so on. It is used for enhancing a desired signal while suppressing noises and interferences at the output of arrays. Large-scale adaptive BF algorithms are often characterized in two important aspects: the convergence rate and the computational complexity. First, the performances of BF algorithms are severely degraded if the weights are incapable of adapting fast enough to the changing interference environment. Typically, the number of homogeneous samples required for the conventional BF to obtain an acceptable performance (within 3dB inferior to the optimum) is about twice the number of the array elements [7].

This work was supported by the National Natural Science Foundation of China under Grant no. 61271293.
[*] E-mail address: xiehumor@gmail.com.

However, the usable samples are very limited for large-scale arrays, especially when the external interferences are space-heterogeneous and time-variable. Second, the high computational complexity also limits the on-line applications of the conventional large-scale adaptive BF. In addition, to obtain high resolution and good performance, the number of modern array sensors [8,9] is so huge that it may be significantly larger than that of usable samples, which may cause the failure of the conventional adaptive BF. Moreover, problems of insufficient samples and heavy computation load are also widely encountered in the multiple-input multiple-output MIMO radar [10] and the space-time adaptive processing (STAP) [11] system whose degrees of freedom are frequently greater than dimensions of the external interference subspace. Therefore, quick convergence and computationally-efficient adaptive BF techniques for large arrays or high-dimension systems are urgently required.

The aim of this paper is to reduce the computation load of beamforming for large-scale array. Inspired by linear kernel methods [12,13] which can be found wide applications in machine learning and pattern recognition, we can use the kernel trick to fast obtain the adaptive BF weight vector via the inverse of the low-dimension Gram matrix instead of the inverse of the high dimension SCM, if the number of antenna array elements is very large compared with that of the training samples. The main contributions of this paper are twofold: 1) a novel strategy using the samples to construct the adaptive weight vector is proposed, which makes the well-known minimum variance distortionless response (MVDR) methods be more useful in the BF of a large-scale array; 2) only a low dimension combinational vector needs to be computed, which results in a fast adaptive BF for a large-scale array and under small sample size.

## II. Structure of Optimal Weight Vector

In this paper, we adapt the standard narrowband beamforming model in which a set of $M+1$ far-field narrowband signals (including a desired signal and the $M$ interference sources) impinge on a uniform linear array with $N$ sensors or array elements. The $N \times 1$ receiving signals vector is given by

$$\begin{aligned} \mathbf{x}(t) &= \mathbf{s}(t) + \mathbf{i}(t) + \mathbf{n}(t) \\ &= a_s(t)\mathbf{s}(\theta_s) + \sum_{m=1}^{M} a_m(t)\mathbf{s}(\theta_m) + \mathbf{n}(t) \end{aligned} \quad (1)$$

where $\mathbf{s}(t), \mathbf{i}(t)$ and $\mathbf{n}(t)$ are the desired signal vector, the interference signal vector and noise vector,

respectively. Here, $a_s(t)$ and $a_m(t), m=1,\cdots,M$ are the desired signal snapshot and interference snapshot, respectively, and $\mathbf{s}(\theta_s)$ and $\mathbf{s}(\theta_m), m=1,\cdots,M$ are the array steering vectors in the direction $\theta_s$ and $\theta_m$.

Firstly, we assume that three stochastic vector sequences $\mathbf{s}(t)$, $\mathbf{i}(t)$ and $\mathbf{n}(t)$ are independent of each other. Secondly, noisy vector sequence $\mathbf{n}(t)$ is assumed to be Gaussian white. Thirdly, we assume that the scale of array is so large that there exists condition $M \ll N$ and $L < N$, where $L$ denotes the number of samples. For example, in a modern large-scale array, $N$ can be over 1000 and $L$ is only in a range of tens to hundreds. In the conventional BF, the adaptive weight vector is derived by maximizing the signal-to-interference-plus-noise ratio (SINR) criterion

$$SINR = \frac{\sigma_s |\mathbf{w}^H \mathbf{s}|^2}{\mathbf{w}^H \mathbf{R}_{i+n} \mathbf{w}} \tag{2}$$

where $\mathbf{R}_{i+n} = E\{(\mathbf{i}(t)+\mathbf{n}(t))(\mathbf{i}(t)+\mathbf{n}(t))^H\}$ is interference-plus-noise covariance matrix (ICM), $\sigma_s = E\{a(t) \cdot a(t)^H\}$ is the signal power, and $\mathbf{s}$ denotes the $N \times 1$ desired signal steering vector. It is well known that the $N \times N$ interference-plus-noise covariance matrix has the following form

$$\mathbf{R}_{i+n} = \mathbf{R}_I + \mathbf{R}_n \tag{3}$$

where $\mathbf{R}_I$ represents the semi-definite positive Hermitian interference covariance matrix, and $\mathbf{R}_n$ is the covariance matrix of the receiving noise vector. Since the noises are white, $\mathbf{R}_n = \sigma^2 \mathbf{I}_N$, where $\sigma^2$ is the noise power level. According to the third assumption, we can deduce that the rank of $\mathbf{R}_I$ satisfies condition $rank(\mathbf{R}_I) = M \ll N$, where $M$ is the dimension of interference subspace.

The optimal adaptive weight vector, $\mathbf{w}_{opt} \in \mathbb{C}^{N \times 1}$ for maximizing the output signal-to-interference-and-noise ratio (SINR) of the beamformer is given within a scale factor by

$$\mathbf{w}_{opt} = \mathbf{R}_{i+n}^{-1} \mathbf{s}. \tag{4}$$

The beamformer (4) is also commonly referred to as the minimum variance distortionless response (MVDR) beamformer.

***Remark 1:*** *In practice, autocorrelation matrix* $\mathbf{R}_{i+n}$ *is approximately estimated by* $L$ *samples* $\mathbf{x}(1),\cdots,\mathbf{x}(L)$, *i.e.*

$$\mathbf{R}_{i+n} \approx \mathbf{R} = \frac{1}{L}\sum_{l=1}^{L}\mathbf{x}(l)\mathbf{x}^{\mathrm{H}}(l) = \frac{1}{L}\mathbf{X}\mathbf{X}^{\mathrm{H}} \in \mathbb{C}^{N\times N}. \quad (5)$$

*where* $\mathbf{X} = [\mathbf{x}(1),\mathbf{x}(1),\cdots,\mathbf{x}(L)]$ *is the data matrix formed by training samples. If the number $L$ of the samples is over that $N$ of array elements, then $\mathbf{R}$ has generally full rank, which is called large samples or moderate samples case in this paper; when $L$ is far greater than $N$, $\mathbf{R}$ can be sufficiently estimated, which is called the full samples case; if $L$ is less than $N$, $\mathbf{R}$ can be underestimated, which is called small samples case that is just considered in this paper. In particular, it is worth mentioning that under small samples case, $\mathbf{R}$ has not full rank and its rank is less than or equal to $L$. It is well known that the underestimated $\mathbf{R}$ is more sensitive to noises and errors than those under the cases of the moderate and full samples.*

***Remark 2:*** *It is seen from (4) and (5) that the computation load of (4) mainly consists of two parts: the estimation of ICM* $\mathbf{R}_{i+n}$ *and the computation of* $\mathbf{R}_{i+n}^{-1}$. *The multiplication and division number is marked as the MDN for short, which is used as an index of the computational complexity in this paper. It is easily shown from matrix theory [14] that the total computation complexity of (4) is at least $N^2L$ and $O(N^3)$ MDN. Consequently, the conventional MVDR Beamformer has the huge computation amount which severely limits its real-time applications in large-scale array signal processing.*

The eigenvalue decomposition of $\mathbf{R}_I$ can be expressed as

$$\mathbf{R}_I = \mathbf{U}_I \mathbf{\Lambda}_I \mathbf{U}_I^{\mathrm{H}} \quad (6)$$

where $\mathbf{\Lambda}_I$ is a $M \times M$ diagonal matrix whose elements are the nonzero positive eigenvalues of $\mathbf{R}_I$, $\mathbf{U}_I$ is the $N \times M$ matrix formed by the eigenvectors corresponding to nonzero eigenvalues. By substituting (6) and (3) into (4) and applying the matrix inversion lemma [14], we have the following formulation

$$\begin{aligned}\mathbf{w}_{opt} &= \mathbf{R}_n^{-1}\mathbf{s} - \mathbf{R}_n^{-1}\mathbf{U}_I\mathbf{\Lambda}_I\left(\mathbf{U}_I^{\mathrm{H}}\mathbf{R}_n\mathbf{U}_I + \mathbf{\Lambda}_I\right)^{-1}\mathbf{U}_I^{\mathrm{H}}\mathbf{R}_n^{-1}\mathbf{s} \\ &= \sigma^{-2}\left(\mathbf{s} - \mathbf{U}_I\mathbf{\Lambda}_I\left(\sigma^2\mathbf{I} + \mathbf{\Lambda}_I\right)^{-1}\mathbf{U}_I^{\mathrm{H}}\mathbf{s}\right) \\ &= [\mathbf{s},\mathbf{U}_I]\mathbf{c}\end{aligned} \quad (7)$$

where $\mathbf{c} = \sigma^{-2}\left[1, -\mathbf{s}^{\mathrm{H}}\mathbf{U}_I\mathbf{\Lambda}_I\left(\sigma^2\mathbf{I} + \mathbf{\Lambda}_I\right)^{-1}\right]^{\mathrm{H}}$ is the $(M+1)\times 1$ coefficient vector. Importantly, relation (7) tells us that the optimal weight vector can be constructed by an $M+1$-dimension subspace spanned by target steering vector and interference eigenvectors.

In general, the signal steering vector may be not orthogonal to the interference subspace which often leads to the computation difficulty. To keep the orthogonality between the signal steering vector and the interference subspace and let $\mathbf{s}^H\mathbf{s}=1$, we should turn (7) into

$$\begin{aligned}\mathbf{w}_{opt} &= \sigma^{-2}\left(\mathbf{s} - \mathbf{U}_I\mathbf{\Lambda}_I\left(\sigma^2\mathbf{I}+\mathbf{\Lambda}_I\right)^{-1}\mathbf{U}_I^H\mathbf{s}\right) \\ &= \sigma^{-2}\left\{\mathbf{s} - \left(\mathbf{I}-\mathbf{ss}^H\right)\mathbf{U}_I\mathbf{\Lambda}_I\left(\sigma^2\mathbf{I}+\mathbf{\Lambda}_I\right)^{-1}\mathbf{U}_I^H\mathbf{s} - \mathbf{s}\left(\mathbf{s}^H\mathbf{U}_I\mathbf{\Lambda}_I\left(\sigma^2\mathbf{I}+\mathbf{\Lambda}_I\right)^{-1}\mathbf{U}_I^H\mathbf{s}\right)\right\} \\ &= \sigma^{-2}\left\{(1-\gamma)\mathbf{s} - \left(\mathbf{I}-\mathbf{ss}^H\right)\mathbf{U}_I\mathbf{\Lambda}_I\left(\sigma^2\mathbf{I}+\mathbf{\Lambda}_I\right)^{-1}\mathbf{U}_I^H\mathbf{s}\right\} \\ &= \left[\mathbf{s}, \mathbf{P}_\mathbf{s}^\perp\mathbf{U}_I\right]\tilde{\mathbf{c}} = \left[\mathbf{s}, \tilde{\mathbf{U}}_I\right]\tilde{\mathbf{c}}\end{aligned} \quad (8)$$

Here $\gamma = \mathbf{s}^H\mathbf{U}_I\mathbf{\Lambda}_I\left(\sigma^2\mathbf{I}+\mathbf{\Lambda}_I\right)^{-1}\mathbf{U}_I^H\mathbf{s}$, and $\tilde{\mathbf{U}}_I = \mathbf{P}_\mathbf{s}^\perp\mathbf{U}_I$ denotes the new interference subspace which is orthogonal to the signal steering vector, where $\mathbf{P}_\mathbf{s}^\perp = \mathbf{I} - \mathbf{ss}^H$ is the orthogonal supplement projection matrix associated with steering vector $\mathbf{s}$. In (8), $\tilde{\mathbf{c}} = \sigma^{-2}\left[(1-\gamma), \mathbf{s}^H\mathbf{U}_I\mathbf{\Lambda}_I\left(\sigma^2\mathbf{I}+\mathbf{\Lambda}_I\right)^{-1}\right]^H$ represents the corresponding coefficient vector. Very interestingly, expression (8) shows that the adaptive weight vector also can be obtained by estimating only the low-dimension interference subspace and the coefficient vector $\tilde{\mathbf{c}}$, which provides a foundation for decreasing the computation load of our beamforming method under small samples case.

## III. Adaptive BF Algorithms Based on Linear Kernel Method

Relation (8) demonstrates that the optimum adaptive weight vector can be spanned by the signal-plus-interference subspace when the receiver noises are Gaussian white. For a large scale array, the dimension of the interference subspace is much lower than that of the optimal weight vector. Similarly to a linear kernel method [12, 13], we can directly use the interference subspace to construct the adaptive weight vector to achieve computational reduction. Under the assumption that the training samples are independent and identically distributed (IID) and mainly composed of interference, it can be seen from (8) that the adaptive weight vector can be very well approximated by a linear combination of the desired signal steering vector and the training samples. It also should be noted that we only focus on the adaptive beamforming problem under small samples case, which means the number of training samples $L$ is smaller than that of array elements, i.e.

$L < N$. Using linear kernel methods [12, 13, 15], we only need to compute a low-dimension vector instead of a high $N$-dimensional adaptive weight vector. By removing the coupling of desired signal steering vector and training samples, imitating (8) and adopting the linear kernel methods, the adaptive weight vector is expressible up to a scale as

$$\mathbf{w} = \mathbf{s} + \mathbf{P}_\mathbf{s}^\perp \mathbf{X}\boldsymbol{\beta} \qquad (9)$$

$\boldsymbol{\beta} \in \mathbb{C}^{L \times 1}$ denotes a lower-dimension combination vector. In addition, according to the famous minimum variance distortionless response (MVDR) criterion, we have

$$\begin{aligned} &\min_{\boldsymbol{\beta}} \mathbf{w}^H \mathbf{R} \mathbf{w} \\ s.t. \quad &\mathbf{w} = \mathbf{s} + \mathbf{P}_\mathbf{s}^\perp \mathbf{X}\boldsymbol{\beta} \\ &\mathbf{w}^H \mathbf{s} = 1 \end{aligned} \qquad (10)$$

The ICM $\mathbf{R}$ in (10) is usually unknown and can be generally estimated by (5). Considering that $\mathbf{w}^H \mathbf{s} = 1$ is automatically satisfied by $\mathbf{w} = \mathbf{s} + \mathbf{P}_\mathbf{s}^\perp \mathbf{X}\boldsymbol{\beta}$, problem (10) can be equivalently transformed into an unconstrained quadratic optimal one

$$\min_{\boldsymbol{\beta}} \left( \mathbf{s} + \mathbf{P}_\mathbf{s}^\perp \mathbf{X}\boldsymbol{\beta} \right)^H \mathbf{R} \left( \mathbf{s} + \mathbf{P}_\mathbf{s}^\perp \mathbf{X}\boldsymbol{\beta} \right). \qquad (11)$$

Insert expression (5) into (11) and let the gradient of the cost function (11) with respect to $\boldsymbol{\beta}$ be equal to zero, then if a constant $1/L$ is ignored, we have the following relation

$$\mathbf{X}^H \mathbf{P}_\mathbf{s}^\perp \mathbf{X}\mathbf{X}^H \mathbf{P}_\mathbf{s}^\perp \mathbf{X}\boldsymbol{\beta} + \mathbf{X}^H \mathbf{P}_\mathbf{s}^\perp \mathbf{X}\mathbf{X}^H \mathbf{s} = \mathbf{X}^H \mathbf{P}_\mathbf{s}^\perp \mathbf{X}(\mathbf{X}^H \mathbf{P}_\mathbf{s}^\perp \mathbf{X}\boldsymbol{\beta} + \mathbf{X}^H \mathbf{s}) = \mathbf{0}. \qquad (12)$$

It is immediately followed by

$$\widehat{\mathbf{R}}^2 \boldsymbol{\beta} + \widehat{\mathbf{R}} \mathbf{X}^H \mathbf{s} = \mathbf{0} \qquad (13)$$

where $\widehat{\mathbf{R}}$ denotes a famous Gram matrix [15] or linear kernel matrix and is computed by

$$\widehat{\mathbf{R}} = \mathbf{X}^H \mathbf{P}_\mathbf{s}^\perp \mathbf{X} = \widehat{\mathbf{X}}^H \widehat{\mathbf{X}} = \in \mathbb{C}^{L \times L} \qquad (14)$$

in which $\widehat{\mathbf{X}} = \mathbf{P}_\mathbf{s}^\perp \mathbf{X}$. It is easily verified that (13) can be directly deduced by the following least squares problems

$$\min \left\| \widehat{\mathbf{R}} \boldsymbol{\beta} + \mathbf{X}^H \mathbf{s} \right\|_2^2. \qquad (15)$$

Noticeably, since $\widehat{\mathbf{R}}$ is at least a semi-definite positive matrix and there is usually $\mathbf{X}^H \mathbf{P}_\mathbf{s}^\perp \mathbf{X} \neq \mathbf{0}$, expression (12) can be intuitively simplified into the following form

$$\widehat{\mathbf{R}}\boldsymbol{\beta} + \mathbf{X}^H \mathbf{s} = 0. \tag{16}$$

According to the linear algebra theory [14], we find that equations (13) and (16) include two cases: 1) $\mathbf{s} \notin \text{span}\{\mathbf{X}\}$, which means that the desired signal is absent in the training data, and $\text{rank}(\widehat{\mathbf{R}}) = \text{rank}(\mathbf{X}^H \mathbf{X}) = L$; 2) $\mathbf{s} \in \text{span}\{\mathbf{X}\}$, which implies that the desired signal is present in the training data, and $L - 1 = \text{rank}(\widehat{\mathbf{R}}) < \text{rank}(\mathbf{X}^H \mathbf{X}) = L$. Very interestingly, equations (13) and (16) preserve the following proposition.

**Proposition 1**: Whether $\mathbf{s} \in \text{span}\{\mathbf{X}\}$ or $\mathbf{s} \notin \text{span}\{\mathbf{X}\}$, the solution of formula (13) is equivalent to that of (16) at the sense of the minimum norm solution.

**Proof**: According to the linear algebra theory [14], the minimum norm solutions of (13) and (16) are respectively described by

$$\boldsymbol{\beta} = -\left(\widehat{\mathbf{R}}^2\right)^\dagger \widehat{\mathbf{R}} \mathbf{X}^H \mathbf{s} \tag{17a}$$

$$\tilde{\boldsymbol{\beta}} = -\widehat{\mathbf{R}}^\dagger \mathbf{X}^H \mathbf{s}. \tag{17b}$$

Let us perform the eigenvalue decomposition (EVD) on $\widehat{\mathbf{R}}$ as $\widehat{\mathbf{R}} = \mathbf{V}\boldsymbol{\Sigma}\mathbf{V}^H$, where $\mathbf{V}$ is the $L \times L$ unitary matrix formed by all the eigenvectors, and the diagonal matrix $\boldsymbol{\Sigma}$ consists of all the eigenvalues arranged in non-increasing order. According to the linear algebra theory [14], $\boldsymbol{\Sigma}$ has the following two forms

$$\boldsymbol{\Sigma} = \text{diag}(\sigma_1, \cdots, \sigma_L) \quad \text{for} \quad \mathbf{s} \notin \text{span}(\mathbf{X}) \tag{18a}$$

or

$$\boldsymbol{\Sigma} = \text{diag}(\sigma_1, \cdots, \sigma_{L-1}, 0) = \begin{bmatrix} \widehat{\boldsymbol{\Sigma}} & \mathbf{0} \\ \mathbf{0} & 0 \end{bmatrix} \quad \text{for} \quad \mathbf{s} \in \text{span}(\mathbf{X}) \tag{18b}$$

where $\widehat{\boldsymbol{\Sigma}} = \text{diag}(\sigma_1, \ldots, \sigma_{L-1}) \in \mathbb{C}^{(L-1) \times (L-1)}$. It is well-known that the Moore-Penrose inverse $\widehat{\mathbf{R}}^\dagger$ of $\widehat{\mathbf{R}}$ is given by

$$\widehat{\mathbf{R}}^\dagger = \mathbf{V}\boldsymbol{\Sigma}^\dagger \mathbf{V}^H. \tag{19}$$

More concretely, (19) has also the following two forms

$$\widehat{\mathbf{R}}^\dagger = \mathbf{V}\boldsymbol{\Sigma}^{-1}\mathbf{V}^H \quad \text{for} \quad \mathbf{s} \notin \text{span}(\mathbf{X}) \tag{20a}$$

or

$$\widehat{\mathbf{R}}^{\dagger} = \mathbf{V} \begin{bmatrix} \widehat{\Sigma}^{-1} & \mathbf{0} \\ \mathbf{0} & 0 \end{bmatrix} \mathbf{V}^H = \widehat{\mathbf{V}} \widehat{\Sigma}^{-1} \widehat{\mathbf{V}}^H \quad \text{for} \quad \mathbf{s} \in \text{span}(\mathbf{X}) \tag{20b}$$

where $\widehat{\mathbf{V}}$ denotes the eigenvector matrix formed by the first $L-1$ nonzero eigenvectors. Therefore, the above analysis shows that we have relations

$$\begin{aligned}
\boldsymbol{\beta} &= -\left(\widehat{\mathbf{R}}^2\right)^{\dagger} \widehat{\mathbf{R}} \mathbf{X}^H \mathbf{s} \\
&= -\left(\mathbf{V}\Sigma\mathbf{V}^H \mathbf{V}\Sigma\mathbf{V}^H\right)^{\dagger} \mathbf{V}\Sigma\mathbf{V}^H \mathbf{X}^H \mathbf{s} \\
&= -\left(\mathbf{V}\Sigma\Sigma\mathbf{V}^H\right)^{\dagger} \mathbf{V}\Sigma\mathbf{V}^H \mathbf{X}^H \mathbf{s} \\
&= -\mathbf{V}\left(\Sigma^2\right)^{\dagger} \mathbf{V}^H \mathbf{V}\Sigma\mathbf{V}^H \mathbf{X}^H \mathbf{s} \\
&= -\mathbf{V}\left(\Sigma^2\right)^{\dagger} \Sigma\mathbf{V}^H \mathbf{X}^H \mathbf{s} \\
&= -\mathbf{V}\Sigma^{\dagger} \mathbf{V}^H \mathbf{X}^H \mathbf{s} \\
&= -\widehat{\mathbf{R}}^{\dagger} \mathbf{X}^H \mathbf{s} = \tilde{\boldsymbol{\beta}}
\end{aligned} \tag{21}$$

The above formula shows that Proposition 1 has been proved.

Substituting relation (21) into expression (9), we derive the adaptive weight vector

$$\mathbf{w} = \mathbf{s} + \mathbf{P}_{\mathbf{s}}^{\perp} \mathbf{X}\boldsymbol{\beta} = \mathbf{s} - \mathbf{P}_{\mathbf{s}}^{\perp} \mathbf{X}\widehat{\mathbf{R}}^{\dagger} \mathbf{X}^H \mathbf{s}. \tag{22}$$

The main computation load of (22) is composed of three parts: the computation of $\widehat{\mathbf{R}}$, the inverse of $\widehat{\mathbf{R}}$ and the some matrix and vector multiplications which require $L^2 N$, $O(L^3)$ and $O(LN)$ MDN, respectively. Comparing (22) with the classical SMI and the Eigenspace based beamformer, our method is much more computationally effective.

In the noiseless case, if the number of training samples is greater than the dimension of interference subspace ($L > M$), we have the following results: 1) when $\mathbf{s} \notin \text{span}\{\mathbf{X}\}$, then $\text{rank}(\widehat{\mathbf{R}}) = \text{rank}(\mathbf{X}^H \mathbf{X}) = M < L$; 2) when $\mathbf{s} \in \text{span}\{\mathbf{X}\}$, then $M = \text{rank}(\widehat{\mathbf{R}}) < \text{rank}(\mathbf{X}^H \mathbf{X}) = M+1$. In the noise case, the smallest eigenvalues of $\widehat{\mathbf{R}}$ are mainly determined by the noises, which may appear to be very small and random. Consequently, the combination vector $\boldsymbol{\beta}$ turns out to be also undetermined and lead to performance degradation. Assuming the number of interferences is known as a prior or can be estimated by semi-heuristic techniques such as the Akaike information criterion (AIC) [16] or minimum description length (MDL) [17], a directly way to prevent $\boldsymbol{\beta}$ from being random is that only the first $M$ column of $\mathbf{V}$ are used to compute $\boldsymbol{\beta}$, namely

$$\boldsymbol{\beta} = -\sum_{i=1}^{M} \frac{\mathbf{v}_i^H \mathbf{X}^H \mathbf{s}}{\sigma_i} \mathbf{v}_i = -(\mathbf{V}\tilde{\boldsymbol{\Sigma}}_M \mathbf{V}^H)\mathbf{X}^H \mathbf{s} \quad (23)$$

where $\tilde{\boldsymbol{\Sigma}}_M = diag[\sigma_1^{-1}, \sigma_2^{-1}, ..., \sigma_m^{-1}, \mathbf{0}]$.

## IV. Experiments

In this section, we present simulation results to compare the proposed method with the conventional beamformers. We assume a large uniform linear array with $N = 400$ Omni-directional sensors spaced half a wavelength apart. The number of Monte Carlos trials per point of the performance curves is 100. In all simulations, the interferences, desired signal and noises are assumed to be independent of each other. The interferences' power levels are 30dB with respect to the noise power level and the direction of arrival (DOA) of the desired signal is $3°$ (note that the direction perpendicular to the linear array is taken as $0°$). Three adaptive beamformers are compared in terms of the averaged output SINR loss: the SMI, Eigenspace based beamformer [18] and the proposed method. The averaged output SINR loss is defined as

$$\rho = \frac{1}{MC} \sum_{m=1}^{MC} \frac{\sigma_s^2 |\mathbf{w}_m^H \mathbf{s}|^2}{(\mathbf{w}_m^H \mathbf{R} \mathbf{w}_m) SINR_{opt}} \quad (23)$$

where $MC$ is Monte Carlos trial number, $SINR_{opt}$ denotes the optimal output SINR derived by using the optimal linear weights for a given interference scenario, and $\mathbf{w}_m$ is the adaptive weights computed via a particular BF method. It can be shown that $SIR_{opt} = \sigma_s^2 \mathbf{s}^H \mathbf{R}_{i+n}^{-1} \mathbf{s}$, where $\mathbf{R}_{i+n}$ is the ideal interference covariance matrix. Note that the original SMI cannot be used for the case $L < N$ since the sample covariance matrix (SCM) is singular. Hence, we adapt the Moore-Penrose inverse of the SCM for SMI in this paper. For our method and the Eigenspace based beamformer, we assume that the number of interferences is known as a prior knowledge in this paper.

In the following simulations, it is assumed that the desired signal is always presented in the training samples. In Figs.1-2, a comparison is shown for the three beamformers when the number of narrow-band interferences is varied ($M$ =3 or 6) and the input SNR=-15dB. We plot the output SINR loss versus the number of training samples for each beamformer in Figs.1-2. The best performance that a particular technique can attain is 0dB and the incoming angles of interferences are indicated in the figures caption. It can be seen from Figs.1-2 that the performance of our method is superior to the SMI and the Eigenspace based beamformer under small samples. Fig. 3 displays the output SINR of relative algorithms versus the input SNR for the fixed training data size $L = 30$. It can be seen from Fig. 3 that the Eigenspace based beamformer is limited to high SNR case. However, our method works well regardless of the input SNR. The beampatterns of the proposed method, LSMI and Eigenspace based beamformer are compared in Fig. 4 for $L = 30$ and SNR=-15dB. The beam pattern of our method has a lower side-lobe than the SMI and Eigenspace-based beamformers. Fig. 5 plots the computation time required by the relative algorithms versus the number of samples. It is apparent from Fig. 5 that our method is much more computationally efficient than the conventional LSMI and Eigenspace based beamformer under small sample support.

## V. Conclusions

In this paper, the adaptive weight vector was represented as a linear combination of the training data and the desired signal steering vector by exploiting the facts that the received samples are mainly composed of interference and desired signal and the adaptive weight vector lies in the signal-plus-interference subspace. Using the well-known linear kernel methods, a fast algorithm has been developed and can be applied in large array beamforming under the small sample support. When the number of samples is much smaller than that of array elements, theoretical analysis and experiments demonstrate that the computational complexity of the proposed method is far smaller than the conventional LSMI and Eigenspace based beamformer.

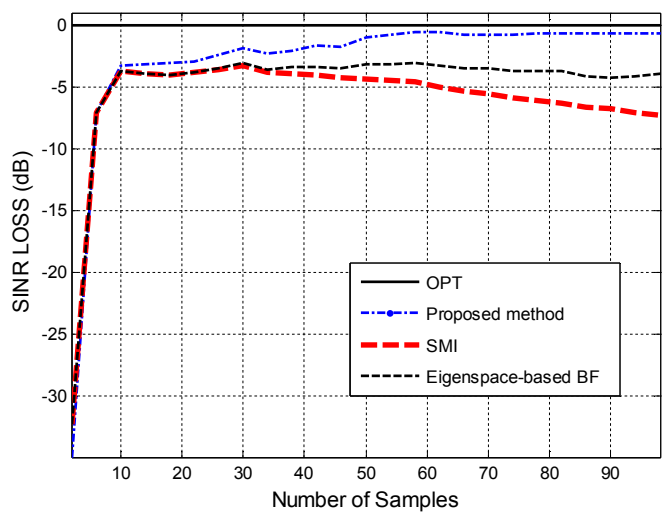

Fig. 1. Averaged output SINR loss versus the number of samples is shown, where the DOA of the interferences are $[-2,\ -4,\ -6]°$.

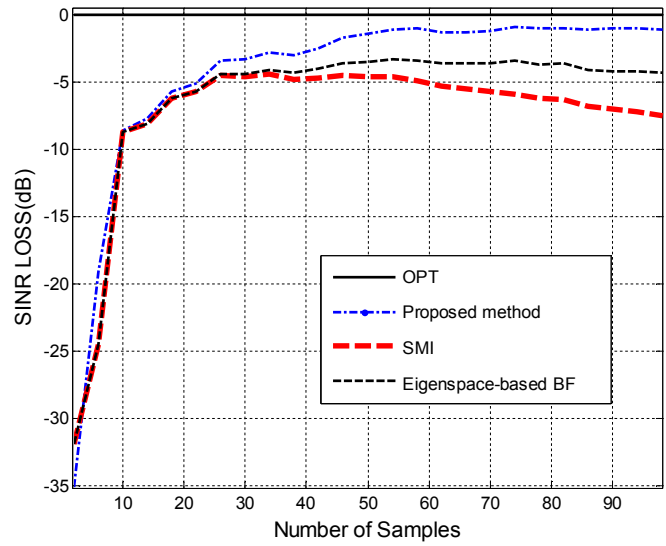

Fig. 2. Averaged output SINR loss versus the number of samples, the DOA of the interferences

are $[-2, -4, -6, 2, 4, 6]°$.

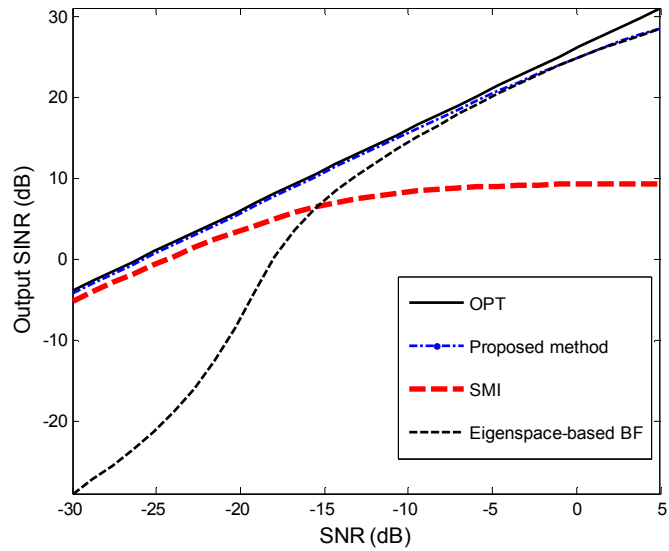

Fig. 3. Output SINR versus SNR, the number of samples L=30.

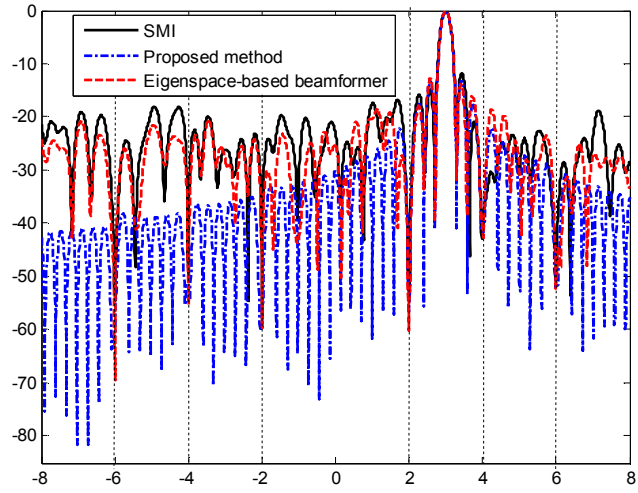

Fig. 4. Beampatterns of the proposed beamformer, LSMI and Eigenspace-based beamformer, the dot vertical lines denote the DOA of interferences. The number of samples is $L = 30$.

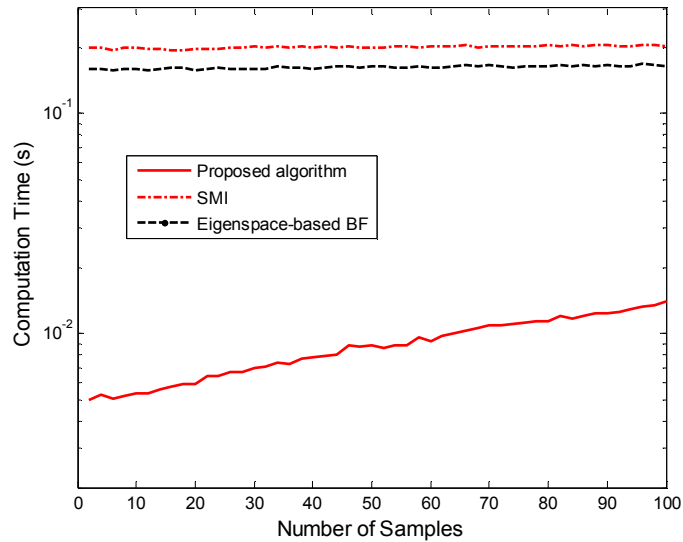

Fig. 5. Computation time versus the number of samples for the three algorithms is shown. Note that the computation time the proposed method takes is much smaller than the calculation time the LSMI and Eigenspace based beamformers spend.

# References

[1] C. Gong, L. Huang, D. Xu and Z. Ye, "Knowledge-aided robust adaptive beamforming with small snapshots," Electronics Letters, vol.49, no.20, pp.1258–1259, 2013.
[2] S. D. Somasundaram, "Wideband robust Capon beamforming for passive sonar," IEEE Journal of oceanic engineering, vol.38, no.2, pp.308-322, Apirl 2013.